\documentclass[apj]{emulateapj}
\usepackage{psfig}

\def\taxp{\hbox{XTE~J1810$-$197}}

\def\s{\phantom 0}

\newcommand\xte{{\it RXTE\/}}

\newcommand\rosat{{\it ROSAT\/}}
\newcommand\chandra{{\it Chandra}}

\newcommand\xmm{{\it XMM-Newton}}

\shorttitle{Spectral Evolution of TAXP XTE J1810--197}
\shortauthors{Gotthelf \& Halpern}

\slugcomment{To Appear in The Astrophysical Journal}
\begin{document}

\title{The Spectral Evolution of Transient Anomalous
X-ray Pulsar XTE J1810--197
} 
\author{E. V. Gotthelf and J. P. Halpern}
\affil{Columbia Astrophysics Laboratory, Columbia
University, 550 West 120$^{th}$ Street,\\ New York, NY 10027-6601;
eric@astro.columbia.edu}

\begin{abstract}

We present a multi-epoch spectral study of the Transient Anomalous
X-ray Pulsar \taxp\ obtained with the {\it Newton X-Ray Multi-Mirror
Mission} (\xmm). Four observations taken over the course of a year
reveal strong spectral evolution as the source fades from outburst.
The origin of this is traced to the individual decay rates of the
pulsar's spectral components. A two-temperature fit at each epoch
requires that the temperatures remains nearly constant at $kT_1 =
0.25$~keV and $kT_2 = 0.67$~keV while the luminosities of these
components decrease exponentially with $\tau_1 = 900$~days and $\tau_2
= 300$~days, respectively.  The integrated outburst energy is $E_1 =
1.3 \times 10^{42}\,d^2_{2.5~kpc}$~ergs and $E_2 = 3.9 \times
10^{42}\,d^2_{2.5~kpc}$~ergs for the two spectral components,
respectively. One possible interpretation of the \xmm\ observations is
that the slowly decaying cooler component is the radiation from a deep
heating event that affected a large fraction of the crust, while the
hotter component is powered by external surface heating at the
foot-points of twisted magnetic field lines, by magnetospheric
currents that are decaying more rapidly.  The energy-dependent pulse
profile of \taxp\ is well modeled at all epochs by the sum of a broad
pulse that dominates in the soft X-rays and a narrower one at higher
energies. These profiles peak at the same phase, suggesting a
concentric emission geometry on the neutron star surface.  The spectral and
pulse evolution together argue against the presence of a significant
``power-law'' contribution to the X-ray spectrum below 8 keV.  The
extrapolated flux is projected to return to the historic quiescent
level, characterized by an even cooler blackbody spectrum, by the year
2007.

\end{abstract}
\keywords{pulsars: general --- stars: individual (XTE J1810--197) --- stars: neutron --- X-rays: stars}

\section{Introduction}

The bright 5.54~s X-ray pulsar \taxp\ is the second example of a
Transient Anomalous X-ray Pulsar (TAXP) and the first one confirmed by
measuring a rapid spin-down rate.  All of its observed and derived
physical parameters are consistent with classification as an Anomalous
X-ray Pulsar (AXP), one that had an impulsive outburst sometime
between 2002 November and 2003 January, when it was discovered
serendipitously by \cite{ibr04} using the {\it Rossi X-ray Timing
Explorer} (\xte).  Its flux was observed to be declining with an
exponential time constant of $269\pm25$~days from a maximum of $F_X
(2-10\ {\rm keV}) \approx 6 \times 10^{-11}$ ergs~cm$^{-2}$~s$^{-1}$.
The source was then localized precisely using two Target of
Opportunity (ToO) observations with the {\it Chandra} X-ray
Observatory by \citet[hereafter Paper~I]{got04} and \citet{isr04}.  In
comparison, archival detections by several X-ray satellites indicate a
long-lived quiescent baseline flux of $F_X(0.5-10\ {\rm keV}) \approx
7 \times 10^{-13}$ ergs~cm$^{-2}$~s$^{-1}$ lasting at least 13~years
and possibly for 23~years prior to the outburst (Paper~I).  Fading of
an IR source within the \chandra\ error circle, similar to ones
associated with other AXPs, confirmed its identification with \taxp\
\citep{rea04a,rea04b}.  The X-ray spectra and pulse profiles from
three observations obtained with \xmm\ during the decline of the
outburst were studied by \citet[hereafter Paper~II]{hal05}.  The short
duty cycle of activity of \taxp\ suggests the existence of a
significant population of as-yet unrecognized, although not
necessarily undetected, young neutron stars (NSs).

In this paper we present a new \xmm\ observation of \taxp.  This data
set, combined with previous \xmm\ observations acquired over the last
year, allows us to characterize the spectral evolution of a Transient AXP.  We
show that the decay rates of the two fitted X-ray spectral components
are distinct, and can be extrapolated in time to approach the previous
quiescent spectrum.  For consistency with Papers I and II, we express
results in terms of a maximum distance of $d=5$~kpc to the pulsar.
However, there is reason to believe that \taxp\ is significantly
closer than this, and we take into account a more realistic estimate
of 2.5~kpc in our discussion of proposed physical models for the X-ray
emission and outburst mechanism.

\section{Observations}

A fourth \xmm\ observation of \taxp\ was obtained on 2004 September
18. The previous three observations are described in Paper~II. We use
the data collected with the European Photon Imaging Camera (EPIC,
\citealt{tur03}) which consist of three CCD imagers, the EPIC~pn and
the two EPIC~MOSs, each sensitive to X-rays in the $0.1-12$~keV energy
range.  In the following we concentrate on data taken with the EPIC~pn
detector, which provided a timing resolution of 48~ms in ``large
window'' mode, for ease of comparison with the earlier data sets. The
fast readout of this instrument ensures that its spectrum is not
affected by photon pileup.  Data collected with the two EPIC~MOS
sensors used the ``small window'' mode and ``timing'' mode, providing
a time resolution of 0.3~s and 1.5~ms, respectively.

\begin{figure}
\centerline{
\psfig{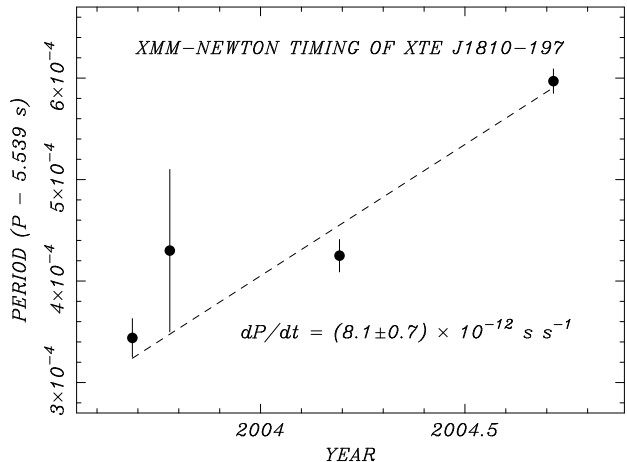}
}
\caption{Period evolution of \taxp\ using timing measurements obtained
with \xmm\ as listed in Table~\ref{summary}.  The dashed line is a
chi-square fit yielding the indicated mean value of $\dot P$. Error
bars are $95\%$ confidence and indicate probable deviations from a
constant spin-down rate.
\label{timing}
}
\end{figure}

For the new data set, we followed the reduction and analysis
procedures used for the previous \xmm\ observations of \taxp, as
outlined in Paper II. The new observation is mostly uncontaminated by
flare events and the filtered data set resulted in a total of 26.5~ks
of good EPIC pn exposure time (24.4~ks live time).  We checked for
timing anomalies that were evident in some previous EPIC~pn data sets
and found none.  Photon arrival times were converted to the solar
system barycenter using the \chandra\ derived source coordinates
R.A. $18^{\rm h}09^{\rm m}51.\!^{\rm s}08$, decl.
$-19^{\circ}43^{\prime}51.\!^{\prime\prime}7$ (J2000.0) given in
Paper~I.

\begin{figure*}
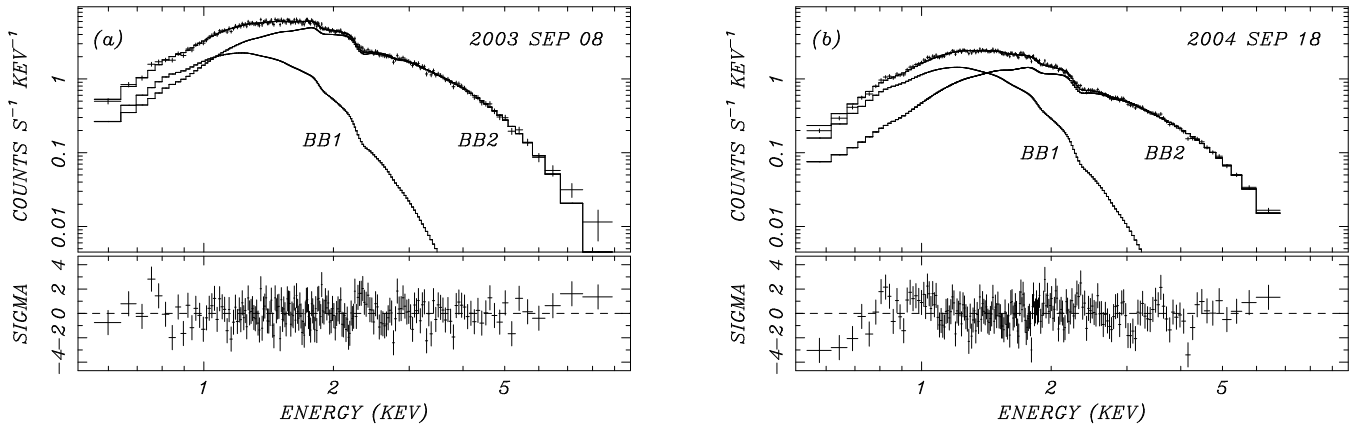

\centerline{
\psfig{figure=f2a.eps,width=0.46\linewidth,angle=270}
\hfill
\psfig{figure=f2b.eps,width=0.46\linewidth,angle=270}
}
\caption{Comparison of \xmm\ EPIC~pn spectra of \taxp\ obtained at
({\it a\/}) the earliest epoch (2003 September) and ({\it b\/}) the
latest epoch (2004 September), fitted with a two-temperature blackbody
model as described in the text and specified in Table~\ref{summary}.
Although the temperatures of the blackbody components have not changed
between the two epochs, the flux of the hot component (BB2) has
decayed faster than the warm one (BB1). Also shown are the residuals
from the best-fit models.
\label{specfig}
}
\end{figure*}

\subsection{Spin-down Evolution of \taxp}

The barycentric pulse period of \taxp\ measured at each \xmm\
observing epoch is given in Table~\ref{summary}. These are derived
from photons obtained with both the EPIC~pn and MOS cameras, with the
exception of the short 2003 October 12 observation for which only
EPIC~pn data of sufficient time resolution is available. The errors
are the 95\% confidence level determined from the $Z_1^2$ test.  As
shown in Figure \ref{timing}, the four period measurements can be
fitted to yield a mean spin-down rate of $\dot P = (8.1 \pm 0.7)
\times 10^{-12}$ s~s$^{-1}$ over the year-long interval. This implies
a characteristic age $\tau_{\rm c} \approx 10,800$~yr, surface
magnetic field $B_{\rm s} \approx 2.1\times 10^{14}$~G, and spin-down
power $\dot E \approx 1.9 \times 10^{33}$ erg~s$^{-1}$, comparable to
the earlier values \citep[and Paper~II]{ibr04}.  Deviations from a
constant $\dot P$ are evident, however, since \citet{ibr04} fitted
values in the range $(1.1-2.2) \times 10^{-11}$~s~s$^{-1}$ in the
first 9 months of the outburst.

As discussed in paper II (\S2.2), timing glitches in AXPs can result
in large increases in their period derivatives [$\Delta \dot P / \dot
P \sim 1$; e.g., 1E~2259+586 \citet*{iwa92,kas03} and 1RXS
J170849.0--400910 \citet{dal03}].  Accordingly, it is possible that
\taxp\ experienced a glitch and an increase in $\dot P$ at the time of
its outburst, and that $\dot P$ is now relaxing to its long-term
value.  But since there is no prior ephemeris for \taxp\ in its
quiescent state, we do not know if its outburst was triggered by a
glitch.  Alternatively, the spin-down torque could have been enhanced
in the early stages of the outburst by an increase over the dipole
value of the magnetic field strength at the speed-of-light cylinder
\citep*{tho02}, or by a particle wind and Alfv\'en waves
\citep*{tho98,har99}, effects that are expected to decline after the
first few months.

\subsection {Spectral Analysis and Results}

\xmm\ observations of \taxp\ have shown that its spectrum is equally
well fitted by a power-law plus blackbody model, as commonly quoted
for AXPs, or a two-temperature blackbody model. In Paper~II we argued
that the two-temperature model is more physically motivated, while the
power-law plus blackbody model suffers from physical inconsistencies.
As we shall show, the new data bolster these arguments, so we
concentrate mainly on the double blackbody model in this work.
Table~\ref{summary} presents a summary of spectral results from all
four \xmm\ observations of \taxp.

\begin{deluxetable*}{lcccc}
\tablecolumns{5} 
\small
\tablewidth{0pt}
\tablecaption{\xmm\ Spectral and Timing Results for \taxp}
\tablehead{
\colhead{Parameter}  &  \colhead{2003 Sep 8} & \colhead{2003 Oct 12} & \colhead{2004 Mar 11}& \colhead{2004 Sep 18}
}
\startdata
$N_{\rm H}$ ($10^{22}$ cm$^{-2}$)\tablenotemark{a}& $0.65 \pm 0.04$      & $0.65 \pm 0.04$       & $0.65 \pm 0.04$     & $0.65$~(fixed)       \\
$kT_1$ (keV)					  & $0.26 \pm 0.02$      & $0.29 \pm 0.04$       & $0.27 \pm 0.02$     & $0.25 \pm 0.01$      \\
$kT_2$ (keV)					  & $0.68 \pm 0.02$      & $0.71 \pm 0.03$       & $0.70 \pm 0.01$     & $0.67 \pm 0.01$      \\
Area $A_1$ (cm$^2$)				  & $1.1\times10^{13}$   & $6.6\times10^{12}$    & $6.8\times10^{12}$  & $9.1\times10^{12}$   \\
Area $A_2$ (cm$^2$) 				  & $6.4\times10^{11}$   & $5.1\times10^{11}$    & $2.9\times10^{11}$  & $2.1\times10^{11}$   \\
BB1 Flux\tablenotemark{b}                         & $4.2\times10^{-12}$  & $5.4\times10^{-12}$	 & $3.5\times10^{-12}$ & $2.6\times10^{-12}$  \\
BB2 Flux\tablenotemark{b}			  & $3.5\times10^{-11}$  & $3.0\times10^{-11}$   & $1.8\times10^{-11}$ & $1.0\times10^{-11}$  \\
Total Flux\tablenotemark{b}			  & $3.93\times10^{-11}$ & $3.84\times10^{-11}$	 & $2.13\times10^{-11}$& $1.29\times10^{-11}$ \\
$L_{\rm BB1}$(bol) (ergs s$^{-1}$)\tablenotemark{c}                & $5.2\times10^{34}$   & $5.1\times10^{34}$ 	 & $3.9\times10^{34}$  & $3.5\times10^{34}$   \\
$L_{\rm BB2}$(bol) (ergs s$^{-1}$)\tablenotemark{c}                & $1.4\times10^{35}$   & $1.3\times10^{35}$ 	 & $7.2\times10^{34}$  & $4.2\times10^{34}$   \\
$\chi^2_{\nu}$(dof)                  		  & 1.1(187)             & 1.1(84)               & 1.1(194)            & 1.2(188)             \\
\\
\tableline
 \\
EPIC pn exposure (ks)                             &     11.5             &     6.9               &     17.0            &     26.5             \\
EPIC pn live time (ks)                            &      8.1             &     6.2               &     15.8            &     24.4             \\
Off-axis angle (arcmin)                           &      0.0             &     8.8               &      0.0            &      0.0             \\
Count rate (s$^{-1}$)\tablenotemark{d}            &     10.6             &     4.8               &      5.8            &      3.4             \\
Epoch (MJD/TDB)\tablenotemark{e}                  &  52890.5642044       &   52924.0000320       &  53075.4999960      &  53266.4999776       \\
Period (s)\tablenotemark{f}                       & 5.539344(19)         & 5.53943(8)            & 5.539425(16)        & 5.539597(12)         \\
\enddata
\tablecomments{\footnotesize Uncertainties in spectral parameters are 90\% confidence for two interesting parameters.} 
\tablenotetext{a}{\footnotesize Parameter derived from a linked fit to all epochs.}
\tablenotetext{b}{\footnotesize Absorbed 0.5--10 keV flux in units of ergs cm$^{-2}$ s$^{-1}$.}
\tablenotetext{c}{\footnotesize Bolometric luminosity assuming $d=5$~kpc.}
\tablenotetext{d}{\footnotesize Background subtracted EPIC pn count rate corrected for detector dead-time.}
\tablenotetext{e}{\footnotesize Epoch of phase zero in Figure~\ref{pulse}.} 
\tablenotetext{f}{\footnotesize Includes EPIC MOS data where available.
95\% confidence uncertainty in parentheses.}
\label{summary}
\end{deluxetable*}

As with the earlier data, the 2004 September source spectrum was
accumulated in a $45^{\prime\prime}$ radius aperture which encloses
$\geq 95\%$ of the energy.  Background was taken from a circle of the
same size displaced $2\farcm3$ along the readout direction.  The
spectra were grouped into bins containing a minimum of 400 counts
(including background) and fitted using the XSPEC package.  For this
analysis the column density is held fixed at $N_{\rm H} = 6.5\times
10^{21}$~cm$^{-2}$, the value determined from previous fits, which are
all consistent.  The best fit to the two-temperature blackbody model
yields temperatures of $kT_1 = 0.25\pm 0.01$~keV and $kT_2 = 0.67\pm
0.01$~keV with a fit statistic of $\chi^2_{\nu} = 1.2$ for 188 degrees
of freedom (Fig.~\ref{specfig}). Each of these two temperatures, which
we refer to as ``warm'' and ``hot'', respectively, remained
essentially the same, nominally to within $4\%$ of that reported for
the first \xmm\ observation (see Table~\ref{summary}). However, the
warm and hot blackbody luminosities declined by 33\% and 70\%,
respectively, in one year.  With the temperature of the hot component
remaining essentially constant, its decline in luminosity is
attributable to a decrease in its emitting area, $A_{2}$.  The most
recent value, $A_2 = 2.1 \times 10^{11}\,d_5^2$~cm$^2$, is $\approx
1\%$ of the NS surface area.  In the case of the warm component, the
decay in flux is modest, so that it is not yet possible to decide
within the errors whether the temperature or the area is the primary
variable.  The most recent area measurement of the warm component,
$A_1 = 9.1 \times 10^{12}\,d_5^2$~cm$^2$, is $\approx 50\%$ of the NS
surface area.

\begin{figure}
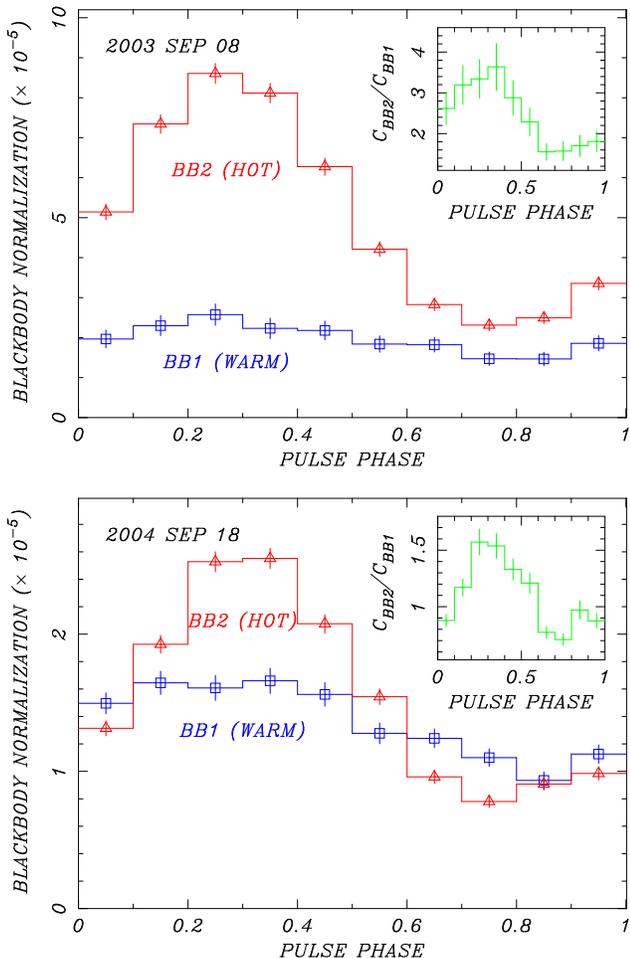

\centerline{
\psfig{figure=f3a.eps,width=0.96\linewidth,angle=-90.}
}
\vskip 0.15in
\centerline{
\psfig{figure=f3b.eps,width=0.96\linewidth,angle=-90.}
}
\caption{ Contribution of the two blackbody components as a function
of rotation phase at two epochs.  ({\it Top}) 2003 September.  ({\it
Bottom}) 2004 September.  The temperatures $kT_1$ and $kT_2$ are held
fixed at the values listed in Table~\ref{summary}, while the
normalization constants are fitted. {\it Inset\/}: For each epoch, the
ratio of blackbody normalization constants as a function of rotation
phase.
\label{phase}
}
\vskip -0.05in
\end{figure}

Phase-resolved spectroscopy using the new \xmm\ data set is compared
here to that reported in Paper II. As before, we fit for the intensity
normalization in each of 10 phase bins with the two temperatures and
column density fixed at the phase-averaged value listed in
Table~\ref{summary}. This is equivalent to determining the relative
projected area of emission as a function of rotation phase. The fits
are again found to be each statistically acceptable as a set, so no
significant test can be made for variations in additional parameters
such as the temperatures or column density. The results of the
phase-resolved spectral fits are shown in
Figure~\ref{phase}. Comparison between the data sets taken a year
apart shows that the modulation with phase of each blackbody component
has remained steady to within the measurement errors.  The phase
alignment of the two temperature components has remained the same, and
the pulses peak at the same phase. This is consistent with the picture
of a small, hot region surrounded by a warm, concentric annulus that
occupies $\sim 1/2$ the surface area of the star.  Neither component
disappears at any rotation phase.  In particular, if the hot component
were completely eclipsed, the spectral decomposition in
Figure~\ref{specfig} indicates that the light curves at $E > 3$~keV
should dip to zero; clearly they do not.

\subsection{Long-term Flux Decay}

With the set of \xmm\ measurements spanning a year, a more accurate
characterization of the X-ray flux decay of \taxp\ is possible than
that reported in the initial \xte\ study of \citet{ibr04}.  The
bolometric flux over time, shown in Figure~\ref{history}, reveals a
new and complex behavior.  While the \xte\ data were consistent with
an exponential decay of time constant $\tau = 269 \pm 25$ days, the
subsequent \xmm\ measurements show that the inferred bolometric
luminosity of the two spectral components are not declining at the
same rate. From the data presented in Table~\ref{summary} we find that
the luminosity of the hotter temperature component falls exponentially
with $\tau_2 = 300$~days, while the warm component decreases with
$\tau_1 = 900$~days, which, although less reliably characterized, is
clearly longer than $\tau_2$.  The shorter time constant is very close
to the one describing the \xte\ data.  Furthermore, the \xte\ spectrum
itself is fitted best by a single temperature blackbody of $kT =
0.7$~keV \citep{rob04}, without a power-law contribution, which is
consistent with our interpretation of the \xmm\ spectrum.  Since \xte\
is sensitive only at energies $>2$~keV where the hotter blackbody
component dominates, the overall X-ray decay from the beginning of the
outburst to the present appears consistent with separate exponential
time constants corresponding to the two distinct thermal-spectrum
components.  We also note that it is not possible to fit an
alternative power-law temporal decay to the hot blackbody flux; such a
model would require a decay index that steepens with time.

As shown in Figure~\ref{history}, we expect that the X-ray flux of
\taxp\ will return to its historic quiescent level by the year 2007.
If the quiescent spectrum is to match that observed before the current
outburst, then it would likely resemble the \rosat\ observation of
1992 March 7.  Although of relatively poor quality, the \rosat\
spectrum can be reasonably well fitted with a single blackbody of $kT
= 0.18\pm 0.02$~keV covering $1.2\times10^{13}\,d_5^2$~cm$^2$ and
$L_{BB}({\rm bol}) = 1.3 \times 10^{34}\,d_5^2$~ergs~s$^{-1}$ (Paper
I).  This blackbody is significantly cooler and larger than the
outburst warm component of $kT_1 = 0.25$~keV and $A_1 =
9.1\times10^{12}\,d_5^2$~cm$^2$, so it should be present even now,
although masked by the fading outburst emission.  We estimate that the
$kT = 0.18$~keV blackbody spectrum measured with \rosat\ would
contribute 30\% of the latest (2004 September) \xmm\ measured flux at
0.5~keV.  Therefore, we expect that such a cooler component will begin
to dominate the soft X-ray spectrum in late 2006, consistent with the
projection of the flux in Figure~\ref{history}.

\subsection{Pulse Shape Evolution}

The energy-resolved folded light curves from the 2003 September 8
\xmm\ observation (Fig.~\ref{pulse}a) show that the pulse peak, in
general, is somewhat narrower than a sinusoid, an effect that is more
pronounced at higher energy. Examination of the light curves measured
one year later (2004 September 18; Fig.~\ref{pulse}d) shows a clear
energy-dependent change.  This is easily seen by forming the ratio of
the curves taken at the two epochs, normalized by their count-weighted
mean.  These ratio curves, presented in Fig.~\ref{ratio}, show changes
in amplitude of up to 20\%.  Specifically, the pulse shape of the
lower-energy light curves has evolved to a smaller amplitude, more
sinusoidal profile. This change is highly significant as determined by
the reduced $\chi^2_\nu$ (24 degrees-of-freedom) for the null
hypothesis of a flat line (see labels on Fig.~\ref{ratio}). In
particular, the most pronounced variation is found in the 1.0--1.5~keV
band with $\chi^2_\nu = 6.3$ corresponding to an infinitesimal
probability of the pulse shape being constant. In contrast, the pulse
shape of the higher energy bins remained essentially unchanged, as
evidence by their $\chi^2_\nu$ values near unity. This trend is
reflected in the two intervening observations (Fig.~\ref{pulse}c and
\ref{pulse}d) as well.

\begin{figure}
\vskip 0.1in
\centerline{
\psfig{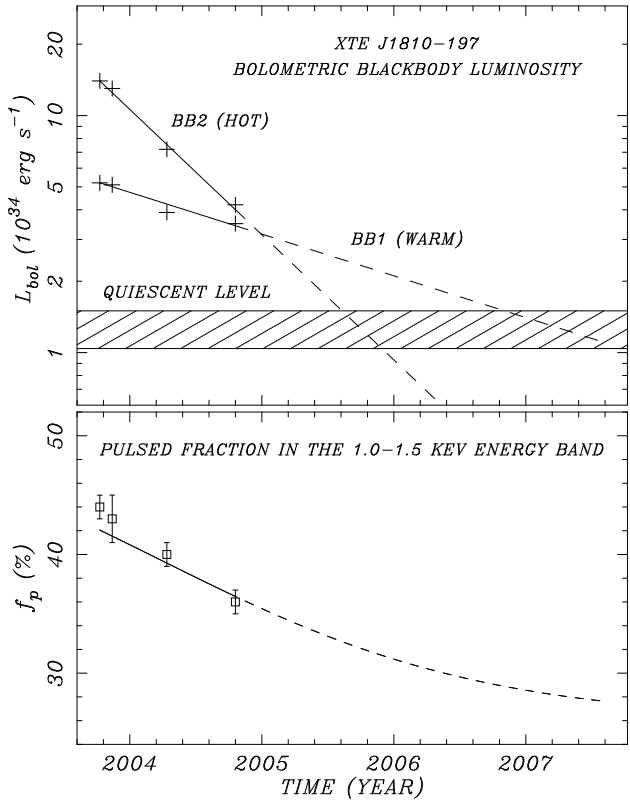}
}
\caption{The bolometric luminosity and pulsed fraction of \taxp\ as a
function of time. ({\it Top\/}) The bolometric luminosity of the
individual components of the two-temperature blackbody model ({\it
crosses\/}) derived from spectral fits to \xmm\ data with the
temperatures fixed (see text). The fitted e-folding times are $\tau_1
= 900$~d and $\tau_2 = 300$~d for the warm and hot blackbody
components, respectively.  The fitted quantities have been
extrapolated to the ($1\sigma$) quiescent range measured in Paper~I
({\it cross-hatched area}).  ({\it Bottom\/}) Pulsed fraction
measurements in the $1.0-1.5$~keV energy band. The {\it solid line}
represents the pulsed fraction modeled in \S 2.4 
from the fitted decay of the blackbody components shown in the top
panel.
\label{history}
}
\end{figure}

The pulse profile must be made up of some combination of pulsed
emission intrinsic to each of the two measured spectral components. In
any given energy band, the changing relative contribution of the soft
and hard spectral components forces the combined pulse shapes to
evolve in time. One way of quantifying this change is to define the
pulsed fraction $f_p$, the fraction of flux above the minimum in the
light curve. This is found to increase smoothly with energy from 34\%
at less than 1~keV to $\approx 51\%$ above 5~keV at the earlier epoch
(see Table~\ref{pulsefit} and Fig.~\ref{pulse}). At the latest epoch
reported herein, the pulsed fraction in the lower energy band
decreased to 26\% while it remained essentially unchanged at the
higher energy.

The simplest hypothesis for modeling the light curve evolution is to
assume that the intrinsic light curve of each spectral
component is steady in time.  If so, the pulsed fraction of the light
curve at a given energy and epoch $f_p(E,t)$ can be expressed as a
linear combination of two light curves of fixed pulsed fractions
$f_{BB1}$ and $f_{BB2}$ weighted by the ratio of counts from the
respective spectral component $N_{BB1}(E,t)$ and $N_{BB2}(E,t)$,

\begin{eqnarray} 
f_p(E,t) = & {{f_{BB1}\,N_{BB1}(E,t)  + f_{BB2}\,N_{BB2}(E,t)} \over {N_{BB1}(E,t)  + N_{BB2}(E,t)}}
\cr\cr
         = & {{f_{BB1}[1-R(E,t)] + f_{BB2}\,R(E,t)}},
\label{pf}
\end{eqnarray}

\medskip 

\noindent where $R(E,t) = N_{BB2}(E,t) / [N_{BB1}(E,t) +
N_{BB2}(E,t)]$ is the normalized flux ratio. This expression is only
valid in the case that the minima of the two light curves coincide, as
is evident for the \taxp\ profiles. For an exponential decay of the
component luminosities, the time-dependent model for the pulsed
fraction is simply,

\begin{equation}
\label{pftime}
 f_p(E,t) = {f_{BB1} + f_{BB2}\, r(E,t_o)\, e^{-(t-t_o)/\tau_d}
 \over 1 + r(E,t_o)\,e^{-(t-t_o)/\tau_d}},
\end{equation}
\medskip 

\noindent where $r(E,t_o) = N_{BB2}(E,t_o) / N_{BB1}(E,t_o)$ is the
flux ratio at time $t_o$ and $\tau_d = (1/\tau_2 - 1/\tau_1)^{-1} =
450$~d is the differential time constant.

We can test this hypothesis directly by fitting for $f_{BB1}$ and
$f_{BB2}$ using a joint least-squares fit to the set of 24 (six energy
bands at four epochs) measured pulsed fractions $f_p(E,t)$ given in
Table~\ref{pulsefit} and the ratio $R(E,t)$ for the double blackbody
spectral model tabulated in Table~\ref{specratio}.  Treating the ratio
$R(E,t)$ as the independent variable, the best fit yields $f_{BB2} =
52 \pm 3\%$ for the hotter component and $f_{BB1} = 26 \pm 2\%$ for
the cooler one. The fit statistic is $\chi^2_{\nu}=1.17$ for 22
degrees-of-freedom, corresponding to a probability $\wp(\geq 1.17) =
0.27$. The lower panel of Fig.~\ref{history} shows the best fit model
for the decay of the the 1.0--1.5~keV pulsed fraction over time and
its extrapolation to an epoch dominated by the warm spectral
component. On the other hand, if we apply the blackbody plus power-law
model, the corresponding values of $R(E,t)$, also listed in
Table~\ref{specratio}, are completely different and the result is
$\chi^2_{\nu}=1.70$ for 22 degrees-of-freedom, corresponding to a
probability $\wp(\geq 1.70) = 0.021$, 13 times less likely relative to
the double blackbody model, for the assumption of fixed intrinsic
pulsed fractions.

Evidently the pulsed fraction of the hotter component is much higher
than for the cooler component.  This difference would account for the
gradual shift from a sharper, more triangular pulse shape found at
higher energies, to a rounder and more symmetric sinusoidal light
curve seen at lower energies.  As was hypothesized above, the decrease
in pulsed fraction with time at low energies follows from the fact
that the hot spectral component is decaying more rapidly than the
cooler one, and so its contribution to the pulsed fraction in the
low-energy bin is declining. To the extent that we are not able to
discern differences greater than $\approx 4\%$, which represents the
measurement uncertainty, the pulsed fractions $f_{BB1}$ and $f_{BB2}$
are individually independent of energy. However, we also cannot rule
out a model in which $f_{BB1}$ itself increases with increasing
energy.

\begin{figure*}
\vskip 0.15in
\epsscale{1.0}
\centerline{
\psfig{figure=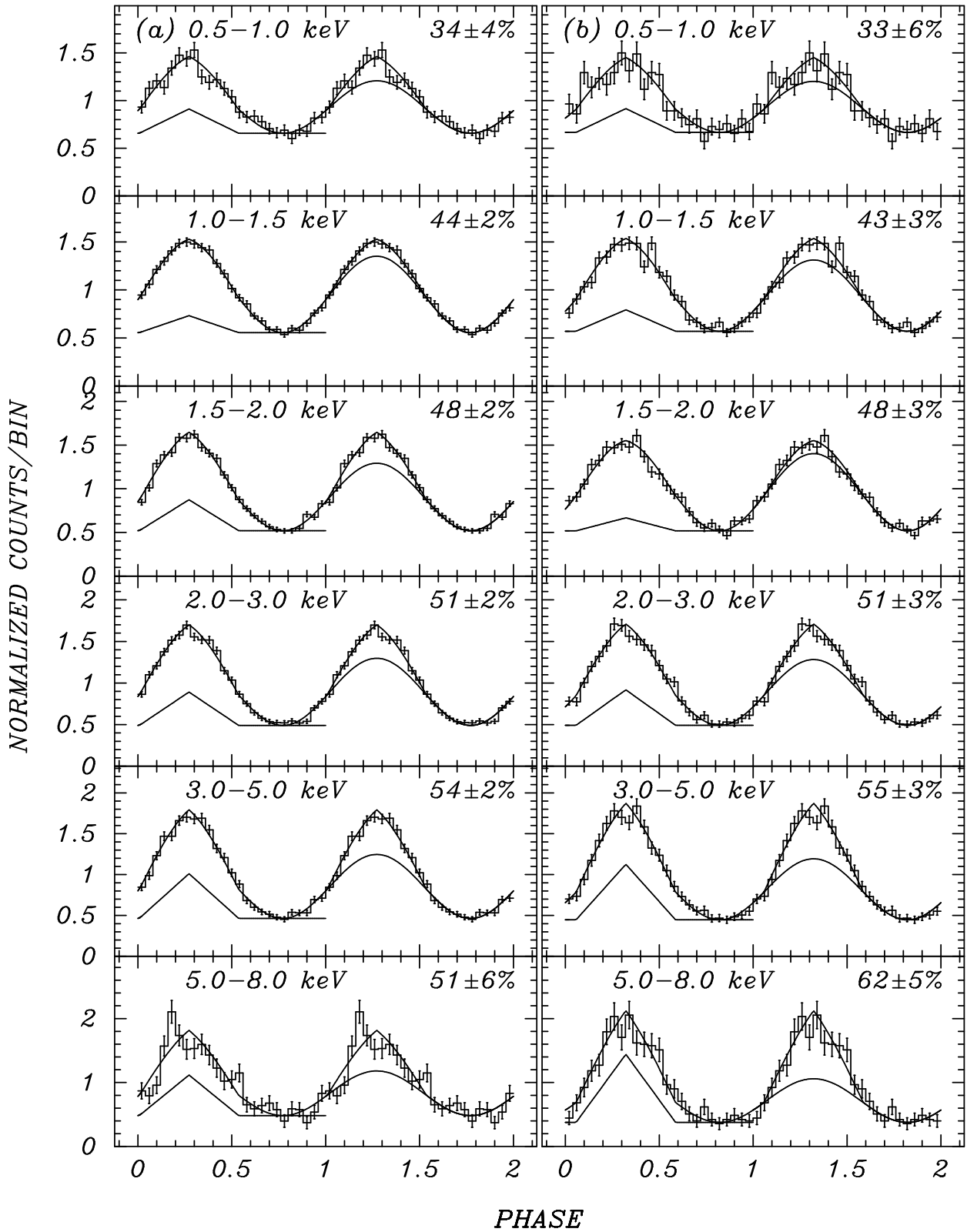,width=0.9\linewidth,angle=270}
}
\vskip 0.15in
\caption{Energy-dependent pulse profiles of \taxp\ obtained with the
\xmm\ EPIC~pn detector. This page shows profile from epochs ({\it
a\/}) 2003 September 8 and ({\it b\/}) 2003 October 12. Also shown is
the best fit to the two-component model for the pulse profile ({\it
solid line\/}) described in the text (see \S 2.4). The individual
contibutions are plotted for the triangular component (Phase $0-1$) and
the sinusoidal component (Phase $1-2$).
\label{pulse}
}
\vskip 0.05in
\end{figure*}

\begin{figure*}
\vskip 0.15in
\epsscale{1.0}
\centerline{
\psfig{figure=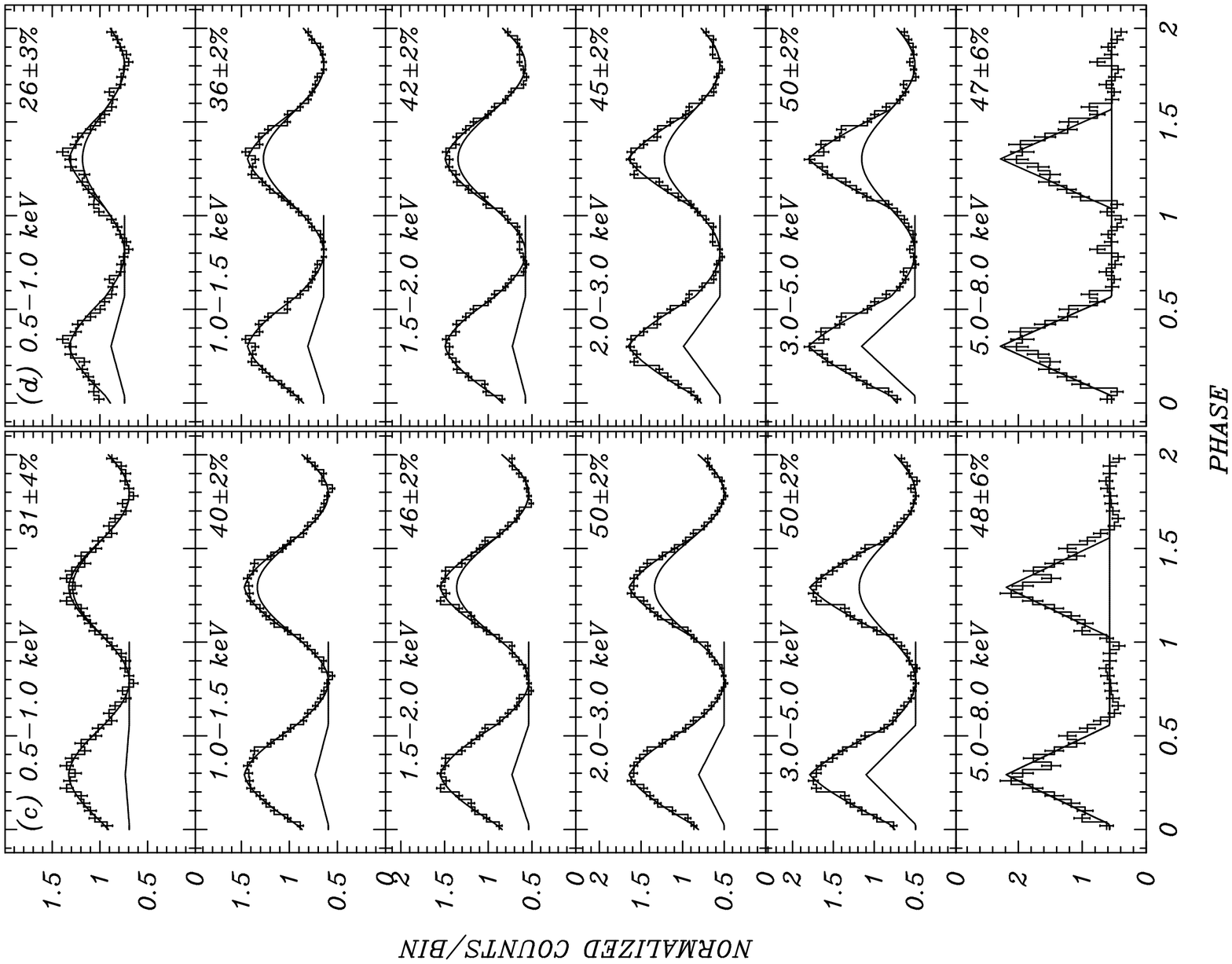,width=0.9\linewidth,angle=270}
}
\vskip 0.15in {Fig. 5. {\it Continued} -- Energy-dependent pulse
profiles of \taxp\ obtained with the \xmm\ EPIC~pn detector for ({\it
c\/}) 2004 March 11 and ({\it d\/}) 2004 September 18.  The epochs of
phase zero are given in Table~\ref{summary}.  Background has been
subtracted. Pulsed fraction at low X-ray energies has decreased in time, while
it has remained essentially unchanged at high energy.
}
\vskip 0.05in
\end{figure*}

It is useful to model the shape of the light curves of \taxp\ in
detail to further quantify their evolution. Overall, the pulse
profiles suggest some linear combination of sinusoidal and triangular
functions, for the soft and hard X-rays, respectively. In the
following, we model the photon counts as a function of phase,
$N(\phi)$, at a given energy $E$ and epoch $t$ by the two-component
model,

$$ N(\phi;E,t) = N_S(\phi;E,t) + N_T(\phi;E,t), $$
where
$$ N_S(\phi;E,t) = \alpha(E,t)\,[\, 1 + \cos (\phi-\phi_S)\,] + \gamma_S(E,t) $$
and

\begin{equation}
\label{profile}
N_T(\phi;E,t) =
\cases{
\beta(E,t)\,[\, 1 - 2|\phi-\phi_T|/\delta(E,t)\,] + \gamma_T(E,t), \cr
\hfill   \quad if |\phi-\phi_T| < \delta/2\cr
\gamma_T(E,t), \hfill if |\phi-\phi_T| \geq \delta/2 \cr 
}
\end{equation}
\medskip

\noindent Here, $\alpha$ is the amplitude of the pulsed signal and
$\gamma_S$ represent the ``unpulsed,'' or minimum level, for the
sinusoidal component, $\beta$ and $\gamma_T$ are the analogous
parameters for the triangular component, and $\delta$ is the duty
cycle (full width) of the triangular pulse.

Initial fits to the light curves indicate that the model can be
constrained by fixing the triangular width $\delta$ to a single value
for all observations. The same proves true for the relative phase
$\phi_S - \phi_T$ between the peak fluxes of the two components.
Accordingly, we used a bootstrap approach to determine these shape
parameters.  After finding a global fit by iteration, the triangular
width is set at $\delta/2\pi = 0.53$ cycles. We then determined
$\phi_S - \phi_T$, which is found to be consistent with zero.
Finally, the absolute phase $\phi_S$ at each epoch was fixed. We
therefore conclude that there is no need to allow energy or time
dependence for the individual component pulse shapes.

\begin{deluxetable}{lcccc}
\tablecolumns{5} 
\small
\tablewidth{0pt} 
\tablecaption{\xmm\ Pulse Profile Fit Results for \taxp}
\tablehead{
\colhead{Energy} &  \multispan2{\hfill 2003\hfill} & \multispan2{\hfill 2004\hfill }\\
\colhead{Range} & \colhead{Sep 8} & \colhead{Oct 12} & \colhead{Mar 11} & \colhead{Sep 18}\\
\colhead{(keV)} & \colhead{(\%)} & \colhead{(\%)} & \colhead{(\%)} & \colhead{(\%)}
}
\startdata
\multispan5{\hfil \hbox{Pulsed Fraction $f_p$ and (Fit Statistic $\chi^2_{\nu}$)\tablenotemark{a}}\hfil \vspace{4pt}}
\\\tableline
$0.5-1.0$ & $34 \pm 4~(0.9)$ & $33 \pm 6~(1.2)$ & $31 \pm 4~(0.5)$ & $26 \pm 3~(0.9)$ \\  
$1.0-1.5$ & $44 \pm 2~(0.6)$ & $43 \pm 3~(1.5)$ & $40 \pm 2~(0.8)$ & $36 \pm 2~(1.2)$ \\
$1.5-2.0$ & $48 \pm 2~(1.5)$ & $48 \pm 3~(1.3)$ & $46 \pm 2~(1.5)$ & $42 \pm 2~(1.6)$ \\
$2.0-3.0$ & $51 \pm 2~(1.4)$ & $51 \pm 3~(1.2)$ & $50 \pm 2~(0.9)$ & $45 \pm 2~(1.9)$ \\
$3.0-5.0$ & $54 \pm 2~(1.2)$ & $55 \pm 3~(1.0)$ & $50 \pm 2~(1.0)$ & $50 \pm 2~(1.5)$ \\
$5.0-8.0$ & $51 \pm 6~(2.0)$ & $62 \pm 5~(0.9)$ & $48 \pm 6~(1.5)$ & $47 \pm 6~(1.6)$ \\
\cutinhead{Normalized Sinusoidal Component Amplitude $\alpha$}
$0.5-1.0$ &  $28 \pm 4$ & $27 \pm 7$ & $30 \pm 4$ & $22 \pm 3$\\   
$1.0-1.5$ &  $40 \pm 2$ & $37 \pm 3$ & $37 \pm 2$ & $31 \pm 2$\\
$1.5-2.0$ &  $39 \pm 2$ & $44 \pm 3$ & $41 \pm 2$ & $39 \pm 2$\\
$2.0-3.0$ &  $40 \pm 2$ & $39 \pm 3$ & $42 \pm 2$ & $33 \pm 2$\\
$3.0-5.0$ &  $39 \pm 3$ & $37 \pm 4$ & $34 \pm 3$ & $33 \pm 3$\\
$5.0-8.0$ &  $35 \pm 8$ & $34 \pm 8$ & $15 \pm 8$ & $06 \pm 8$\\
\cutinhead{Normalized Triangular Component Amplitude $\beta \delta/2$}
$0.5-1.0$  & $\s6 \pm 3$ &  $\s6 \pm 4$ & $\s1 \pm 2$ & $\s4 \pm 7$\\   
$1.0-1.5$  & $\s5 \pm 2$ &  $\s6 \pm 2$ & $\s4 \pm 1$ & $\s4 \pm 6$\\ 
$1.5-2.0$  & $10 \pm 2$ &  $\s4 \pm 2$ & $\s5 \pm 1$ & $\s4 \pm 6$\\ 
$2.0-3.0$  & $11 \pm 2$ &  $11 \pm 3$ & $\s8 \pm 1$ & $11 \pm 6$\\ 
$3.0-5.0$  & $14 \pm 2$ &  $18 \pm 3$ & $16 \pm 2$ & $17 \pm 5$\\ 
$5.0-8.0$  & $14 \pm 5$ &  $27 \pm 6$ & $32 \pm 5$ & $40 \pm 5$\\ 
\cutinhead{Normalized Unpulsed  Amplitude $\gamma$}
$0.5-1.0$ &  $67 \pm 4$ & $66 \pm 3$ & $69 \pm 3$ & $74 \pm 2$\\
$1.0-1.5$ &  $57 \pm 2$ & $56 \pm 2$ & $59 \pm 2$ & $64 \pm 1$\\
$1.5-2.0$ &  $52 \pm 2$ & $52 \pm 2$ & $54 \pm 2$ & $58 \pm 2$\\
$2.0-3.0$ &  $49 \pm 2$ & $49 \pm 2$ & $50 \pm 2$ & $55 \pm 2$\\
$3.0-5.0$ &  $45 \pm 2$ & $46 \pm 2$ & $49 \pm 2$ & $50 \pm 2$\\
$5.0-8.0$ &  $48 \pm 5$ & $38 \pm 4$ & $52 \pm 5$ & $52 \pm 4$\\
\enddata
\tablecomments{Results of fits to the two-component sinusoidal plus triangular model for the pulse profile
described in \S2.4. The fixed model parameters are $\phi_S - \phi_T =
0.0$, $\delta/2\pi = 0.53$~cycle, and at each epoch, the absolute phase of the
sinusoidal function. }
\tablenotetext{a}{Uncertainties in the pulsed fractions are at the
1-$\sigma$ (68\% confidence level) for three interesting parameters
($\alpha$, $\beta$, $\gamma$). }
\label{pulsefit}
\end{deluxetable}

\begin{deluxetable}{lcccc}
\tablecolumns{5} 
\small
\tablewidth{0pt} 
\tablecaption{Component Flux Fraction for \taxp}
\tablehead{
\colhead{Energy} & \multispan2{\hfill 2003\hfill } & \multispan2{\hfill 2004\hfill }\\
\colhead{ }      & \multispan4{--------------------- \hfill  ---------------------}\\
\colhead{Range (keV)} & \colhead{Sep 8} & \colhead{Oct 12} & \colhead{Mar 11} & \colhead{Sep 18}
}
\startdata
\multispan5{\hfil \hbox{$R(E,t) \equiv N_{BB2}/(N_{BB1}+N_{BB2})\tablenotemark{a}$}
\hfil \vspace{4pt}}
\\\tableline
$0.5-1.0$ & 0.389 & 0.332 & 0.285 & 0.218 \\
$1.0-1.5$ & 0.591 & 0.499 & 0.466 & 0.399 \\
$1.5-2.0$ & 0.800 & 0.702 & 0.696 & 0.661 \\
$2.0-3.0$ & 0.941 & 0.886 & 0.896 & 0.893 \\
$3.0-5.0$ & 0.996 & 0.988 & 0.991 & 0.992 \\
$5.0-8.0$ & 1.000 & 1.000 & 1.000 & 1.000 \\
\cutinhead{$R(E,t) \equiv N_{PL}/(N_{PL}+N_{BB})\tablenotemark{b}$}
$0.5-1.0$ & 0.928 & 0.911 & 0.925 & 0.952 \\
$1.0-1.5$ & 0.755 & 0.718 & 0.753 & 0.807 \\
$1.5-2.0$ & 0.533 & 0.480 & 0.524 & 0.567 \\
$2.0-3.0$ & 0.347 & 0.291 & 0.331 & 0.340 \\
$3.0-5.0$ & 0.247 & 0.191 & 0.223 & 0.200 \\
$5.0-8.0$ & 0.376 & 0.278 & 0.325 & 0.261 \\
\enddata 
\tablenotetext{a}{$R(E,t) \equiv N_{BB2}/(N_{BB1}+N_{BB2})$ is the 
fraction of counts  in a given energy band from the hotter component of the two-temperature
blackbody spectrum based on the spectral model presented in Table~\ref{summary}.}
\tablenotetext{b}{$R(E,t) \equiv N_{PL}/(N_{PL}+N_{BB})$ is the fraction of counts in a
given energy band from the power-law component of the best fit
power-law plus blackbody model blackbody spectrum (not shown).}
\label{specratio}
\end{deluxetable}

With the shape of the pulse components fixed, we fit the restricted
model with just $\alpha, \beta$, and $\gamma$ ($ = \gamma_S +
\gamma_T$, since the $\gamma$'s are indistinguishable in this fit) as
free parameters.  This model fitted to all 24 individual pulse
profiles yields a reduced $\chi^2_{\nu}$ ranging from $0.6-2.0$ with a
typical value of $\chi^2_{\nu}=1.2$, corresponding to a probability of
$\wp(\gtrsim \chi^2_{\nu}) = 0.23$ for 22 degrees-of-freedom.  The
best-fit model parameters, derived pulsed fractions, and fit statistic
for each light curve are presented in Table~\ref{pulsefit}. A clear
trend is seen as the pulsed fraction decreases with time.  The
best-fit models are shown in Figure~\ref{pulse} overlaid on plots of
the energy-resolved pulsed profiles for the four epochs.  We checked
that a single component model, either a pure sinusoidal or triangular
pulse, does not adequately characterize the data based on an
unacceptably large $\chi^2_{\nu}$ for many of the fits.

It is possible to further reduce the number of free parameters in the
fit by constraining the coefficients $\alpha, \beta$, and $\gamma$ so
that the phase-averaged photon count ratios $N_S(E,t)/N_T(E,t)$ are
identical with the ratios derived from the two-temperature blackbody
spectral fits in each energy interval at each epoch.  That is, we test
the hypothesis that the warm blackbody is responsible for only the
sinusoidal pulse and the hot blackbody for only the triangular pulse
by requiring $N_S(E,t)/N_T(E,t) = N_{BB1}(E,t)/N_{BB2}(E,t)$ at all
epochs.  We find that, while reasonable fits to the light curves are
possible for the lower and highest energy bands, fits in the
$1.5-5.0$~keV range proved unacceptable. This argues against a simple
one-to-one correspondence between the spectral and the chosen light
curve component decomposition. It is possible that we have not started
with the correct spectral or light curve basis pairs or that the basic
pulse profiles are each energy dependent.  It is also possible that
there may be a third, unfitted softer component associated with the
quiescent flux, as seen by ROSAT (see \S 2.3).  Continuing observation
of the spectrum as the source fades should clarify the relationship
between spectral and light curve components.

\section{Interpretation}

\subsection{Spectral Evolution: Constraining Models}

While the $0.5 - 8$~keV X-ray spectra of AXPs (including \taxp)
clearly require a two-component fit, it is not possible to prove from
spectra alone that either of two very different models, namely a
power-law plus blackbody, or a two-temperature blackbody, is
correct. For \taxp, either model yields an acceptable chi-square when
fitted to the X-ray spectrum.  However, in Paper~II we presented
physical arguments against the properties of the particular power law
that results from the power-law plus blackbody fit.  Unlike some
rotation-powered pulsars whose spectra are fitted by hard power laws
plus soft blackbodies, in AXPs the roles of these two components are
reversed. For \taxp, most of the X-rays belonging to the $\Gamma
\approx 3.7$ power-law component are {\it lower} in energy than those
fitted by the blackbody (see Figure 5a of Paper II). This steep power
law must turn down sharply just below the X-ray band in order not to
exceed the faint, unrelated IR fluxes. As shown in Paper II, this
would be difficult to achieve in a synchrotron model.  And in a
Comptonization model, where are the seed photons?  Even if such a
sharp cutoff existed at the low end of the EPIC energy band, a model
in which the power-law component is due to Compton upscattering of
thermal X-rays from the surface does not explain why most of the
power-law photons have lower energy than the observed blackbody
photons unless there is a larger source of seed photons at energies
below 0.5~keV.  However, such a seed source would have to be thermal
emission from the neutron star.  Since the two-temperature model uses
a large fraction of the surface area for the $0.5-2$~keV photons, we
prefer to regard the X-rays in this band as plausible seeds for,
rather than as the product of, inverse Compton scattering. Supporting
this assumption in the specific case of \taxp\ are the shapes of its
pulse profiles. The smooth increase in pulsed fraction with energy,
and the strict phase alignment at all energies, are unnatural under
the hypothesis of two different emission mechanisms and locations.

The evolving spectral shape and pulse profiles of \taxp\ during its
decline from outburst offer new and complementary evidence concerning
the appropriate spectral decomposition. We find that the the pulsed
fraction is declining with time at low energies, while remaining
essentially constant at high energies.  This fact is consistent with
the changing contributions of fluxes in the two-temperature fit as the
two components decline at different rates.  The warm and hot
components contribute significantly to the soft X-rays, whereas the
hard X-rays are supplied entirely by the hot component.  This is
quantified in Table~\ref{specratio}, where it is seen that the hot
blackbody component accounts for 30--60\% of the flux below 1.5 keV,
and $>99\%$ of the flux above 3~keV at all times.  Under this spectral
decomposition, the pulsed fraction at energies above 3~keV is large
and unchanging because one and only one spectral component is present
at those energies, and it has the larger pulsed fraction of the two
spectral components.  In contrast, the pulsed fraction at low energies
($<1.5$ keV) is understood to decrease in time because the hot
component makes only a fractional contribution to its light curve, and
its fraction decreases faster than the warm spectral component, which
has the intrinsically smaller pulsed fraction.  This effect is also
illustrated in Figure~\ref{history}, where the modeled pulse fraction
is compared to the data.

\begin{figure}
\vskip 0.15in
\centerline{
\psfig{figure=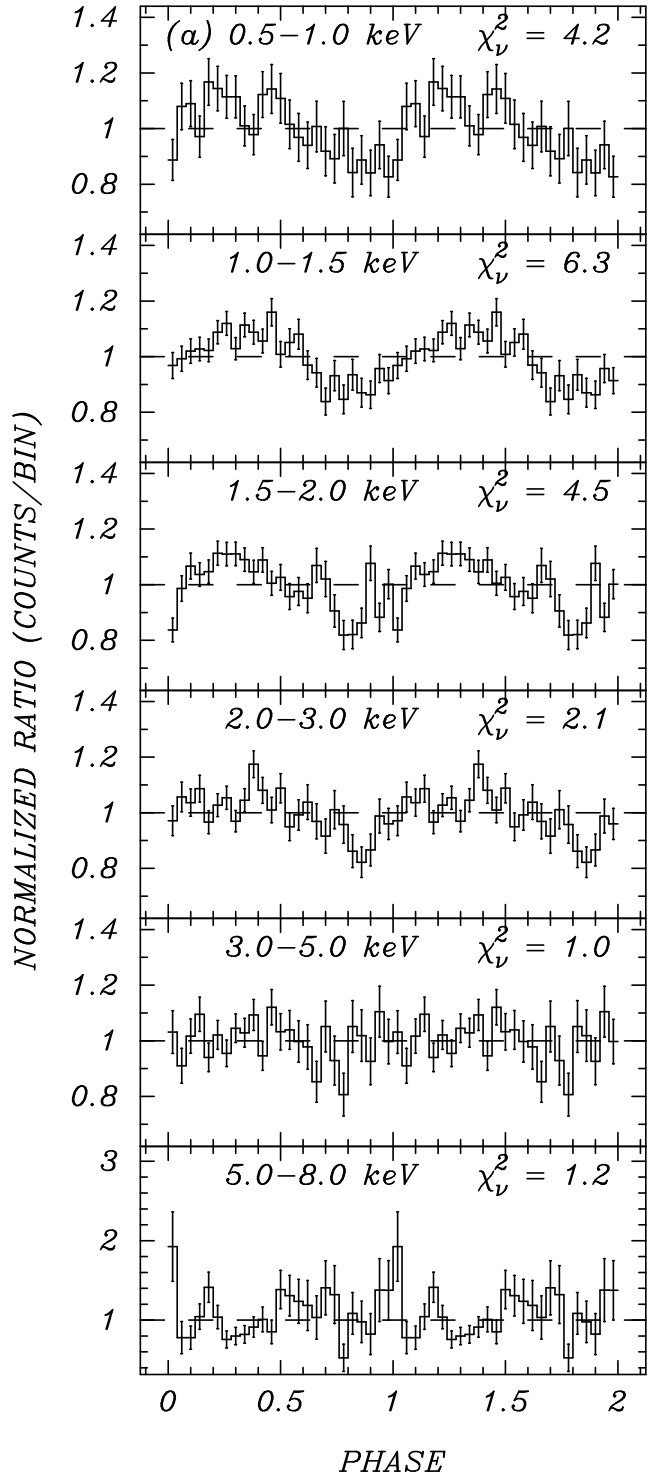,width=0.95\linewidth,angle=270}
}
\vskip 0.15in
\caption{ Comparison of the \xmm\ EPIC~pn pulse profiles of \taxp\ obtained 
one year apart in six energy bands.  Displayed is the ratio
in each band between the 2003 September 8 and 2004 September 18
profiles shown in Fig.~5, normalized by the count-weighted mean. The
profiles show more systematic change at lower energies as indicated by
the ratio of the amplitudes (vertical scale). 
\label{ratio}
}
\vskip 0.05in
\end{figure}

In contrast, consider how the spectral components would contribute to
the various energy bands in the power-law plus blackbody spectral
model.  Table 3 shows that, unlike the two blackbody model, a fitted
power-law component accounts for 79\% of the soft ($<1.5$~keV) X-rays
at {\it all\/} epochs, while decreasing its contribution to the hard
($> 3$~keV) X-rays over time. In this model, the pulsed fraction
should remain constant at low energies because only the power law is
present there, while at high energies the pulsed fraction should vary
because both the blackbody and the power-law components are present,
in varying proportions.  This spectral decomposition would drive an
evolution of the pulse shapes that is opposite of what is observed.
Thus, we find that the detailed evolution of the X-ray emission from
\taxp\ further supports the assumption of a purely thermal surface
emission model, and leaves no evidence of a steep power law in the
$0.5-8$~keV band, as is commonly fitted to individual observations of
AXPs.

Our decomposition of the pulsed light curves is also consistent with
the absence of modulation in the \rosat\ observation of 1992 March 7,
to an observed upper limit of 24\% in the $0.2-2.0$~ keV PSPC energy
band. Since the \rosat\ spectrum is fitted by an even cooler blackbody
of $kT = 0.18$~keV, it can be expected that the pulsed fraction in
quiescence will fall below the 25\% value fitted to the warm blackbody
in \xmm\ data once that component fades away.

\subsection{Re-evaluation of the Distance to \taxp\ }

For consistency with Papers I and II, we have parameterized results
here in terms of a distance of 5~kpc.  However, this was considered an
upper limit based on an \ion{H}{1} absorption kinematic distance to
the neighboring supernova remnant G11.2$-$0.3 \citep{gre88} and its
X-ray measured column density $N_{\rm H} \sim 1.4 \times
10^{22}$~cm$^{-2}$ \citep{vas96}.  In comparison, the column density
fitted to a power-law plus blackbody model for \taxp\ was $N_{\rm H} =
1.02 \times 10^{22}$~cm$^{-2}$.  Since on physical grounds we strongly
prefer the double blackbody model, which requires $N_{\rm H} = 0.65
\times 10^{22}$~cm$^{-2}$, it seems appropriate to reduce the distance
estimate to one that is more compatible with the smaller $N_{\rm H}$
value.  We now assume that the distance to \taxp\ is 2.5~kpc, and
explore the consequences of this revision.  The effect is to reduce
the inferred X-ray luminosities and blackbody surface areas.  For $d =
2.5$~kpc, the surface area emitting the warm blackbody component is
$A_1 < 3 \times 10^{12}$~cm$^{-2}$ at all times, which is less than
1/6 the area of a neutron star.  This makes it easier to understand
how the pulsed fraction of the softest X-rays can be as large as
$26\%$.

We can also revise the estimate of the total energy emitted in the
outburst. At a distance of 2.5~kpc, the bolometric luminosity quoted
in Paper II is $L_{\rm BB2} = 1.5 \times
10^{35}\,d_{2.5}^2\,e^{-(t-t_0)/300\,{\rm days}} $ ergs~s$^{-1}$ with
respect to the initial time $t_0$ of the outburst observed by \xte\
\citep{ibr04}, which corresponds to an extrapolated energy of $3.9
\times 10^{42}\,d_{2.5}^2$ ergs.  Since \xte\ is only sensitive to the
hotter blackbody component, we should add another contribution from
the \xmm\ measured warm component, which can be estimated as $L_{\rm
BB1} = 1.7 \times 10^{34}\,d_{2.5}^2\,e^{-(t-t_0)/900\,{\rm days}}$
ergs~s$^{-1}$.  The extrapolation required to integrate the
contribution of the warm component is less certain than for the hot
component, but it corresponds to an energy of only $1.3 \times
10^{42}\,d_{2.5}^2$ ergs.  The total estimated energy is then $5.2
\times 10^{42}\,d_{2.5}^2$ ergs, which is comparable to the amount of
heat assumed to be deposited in the crust during a deep heating event
\citep*{lyu02}, or the extra energy stored in an azimuthally twisted
external magnetic field \citep{tho02}.  In the following section, we
discuss how the observations might probe the actual mechanism of the
outburst.

\subsection{Heating or Cooling?}

Two mechanisms have been invoked to explain the late-time
``afterglow'' of an outburst of a magnetar.  The first involves
surface heating by long-lived currents flowing on closed but
azimuthally twisted external magnetic field lines, and Comptonization
of the resulting surface X-ray emission by the same particles
\citep{tho02}.  The originating event could be a sudden fracture in
the crust in response to twisting of the internal magnetic field, or a
plastic deformation of the crust that gradually transfers internal
magnetic twist to the external field.  It is thought that the major
outbursts of SGRs are magnetically trapped fireballs triggered by a
sudden fracture, but this is not known to apply to \taxp\ because no
bursting behavior was seen, and no prior rotational ephemeris exists
to test for a glitch.  In any case, the extra energy above that of a
pure dipole, $\Delta E$, stored in a twisted magnetic field external
to the star is,

\begin{equation}
\label{benergy}
\Delta E = 1.4 \times 10^{44}\,\Delta\phi^2_{\rm N-S}\,
\left(B_{\rm p} \over 10^{14}\,{\rm G}\right)^2\,
\left(R_{\rm NS} \over 10\,{\rm km}\right)^3\ {\rm ergs},
\end{equation}
\medskip

\noindent where $\Delta\phi_{\rm N-S} < 1$~rad is the azimuthal twist from north
to south hemisphere \citep{tho02}.  This is enough to account for the
extrapolated $5.2 \times 10^{42}\,d_{2.5}^2$ ergs radiated if
$\Delta\phi_{\rm N-S}$ is a small fraction of a radian.  

In this model, the X-ray spectrum is a combination of surface thermal
emission, and Compton scattering and cyclotron resonant scattering of
the thermal X-rays. Compton upscattered flux may result in power-law
emission that dominates at high energies such as that reported above
10~keV for the AXP in Kes~73 \citep{kui04}. However, the flat
power-law index ($\Gamma \simeq 0.94$) does not contribute to the
softer spectrum, in particular below 2~keV. So we consider that the
potential power-law tail in the fainter \taxp\ may no be detectable
yet. The decay timescale is determined by the rate at which power is
consumed by the magnetospheric currents, which is comparable to that
needed to power the observed X-ray luminosities of AXPs in general and
\taxp\ in outburst.  Furthermore, it is the evolution of surface
heating that basically determines the instantaneous X-ray luminosity
and its decay, but more specific predictions of decay curves have not
been made.

In the second picture, a deep crustal heating event, heat deposited
suddenly by rearrangement of the magnetic field is gradually
transferred by diffusion and radiation, and the radiated X-rays are
purely thermal.  While the originating event might have been a
fracture of the crust as in the first scenario, an assumption that the
heat is evenly distributed throughout the crust results in a
particular decay curve from one-dimensional models of heat transfer
\citep{lyu02}.  In this model, the heating is virtually instantaneous,
within $10^4$~s, while it is the rate of conduction cooling that
determines the observed decay curve.  \citet{lyu02} assumed a
deposition of $\sim 10^{25}$ ergs~cm$^{-3}$ to a depth of 500~m, which
is $\leq 1\%$ of the magnetic energy density at the surface.  If
occurring over the entire crust, this is $\sim 1 \times 10^{43}$ ergs,
comparable to the total X-ray energy emitted during the decay of
\taxp.  However, \citet{lyu02} showed that the resulting cooling
luminosity would follow a $t^{-0.7}$ power law, which is rather slow
compared to the observed exponential decay of the hot component of
\taxp, while 80\% of the energy should be conducted into the neutron
star core rather than radiated from the surface.

It is not obvious if the behavior of \taxp\ supports or excludes
either of these models.  One interesting possibility is that the
slowly decaying warm component is the radiation from a deep heating
event that affected a large fraction of the crust, while the hot
component of the spectrum is powered by external surface heating at
the foot-points of twisted magnetic field lines by magnetospheric
currents that are decaying more rapidly.  It is not yet clear if the
flux decay of the warm component is primarily an effect of decreasing
temperature or decreasing area, because the decay has so far been
modest.  It is possible that the decay of the warm component {\it is}
consistent with a $t^{-0.7}$ power law rather than an exponential of
$\tau_1 = 900$~days. Fits to $L_{BB1}$ as a function of time allows a
power-law decay index in the range $-0.3$ to $-0.7$. In contrast,
because of its more rapid decay, it is evident that the hotter
component is declining exponentially, in area rather than in
temperature.  This shrinking in area might be easier to understand as
a decay of the currents or a rearrangement of the magnetic field lines
that are channeling the heating on the surface.  But there is no good
evidence yet of an inverse Compton scattered component from the
particles responsible for that current, at least not at energies
$<8$~keV.  Perhaps the solid angle subtended by the twisted field
lines in the magnetosphere is too small to scatter most of the thermal
photons.  Each of the models is challenged in some aspect by the
observations, but each may still have some applicability. Observing
the decay through to quiescence should help to clarify the situation.

\subsection{Constraints on Emission and Viewing Geometry}

The pulsed light curves offer an additional diagnostic of the emission
and viewing geometry.  If we adopt the hypothesis that virtually all
of the $0.5-8$~keV flux is surface thermal emission, then the symmetry
of the light curves and their strict pulse-phase alignment as a
function of energy argue for a concentric geometry in which a small
hot spot is surrounded by a larger, warmer region.  The large observed
pulsed fractions are achievable in a realistically modeled NS
atmosphere that accounts for the different opacities of the normal
modes of polarization in a strong magnetic field
\citep*{oze01a,oze01b}.

Many geometries were modeled by \citet{oze01a}, but only total pulsed
fractions were reported rather than detailed light curves.  It is
important to model the pulse profiles completely, since the number of
peaks and their phase relationship depend in detail on the surface and
viewing geometry.  In addition, the relative effects of fan versus
pencil beaming, as regulated by the anisotropic opacity in a strong
magnetic field, allow the number of peaks in the light curve to differ
from the number of hot spots on the surface.  Whereas the pulsed
fraction of the hot component in \taxp\ is 53\%, this is not high
enough to require a single spot.  It is possible that the flat
interpulse region of the triangular, hard X-ray pulse is actually
emission from an antipodal spot at large viewing angle.  \cite{oze01b}
showed that for large magnetic fields, most of the flux would be in a
broad fan beam, while \cite{tho02} argue that cyclotron resonant
scattering in the magnetosphere can significantly reduce this effect.
Detailed physical modeling of the pulse profiles of \taxp\ is needed
before definite conclusions about the emission and viewing geometries
can be drawn.

\section{Conclusions and Future Work}

Four sets of \xmm\ observations of \taxp\ spanning a year, clarify
several new behaviors that might apply to other AXPs and AXP-like
objects: 

\begin{itemize}

\item a two-temperature blackbody model is arguably preferred
over the standard blackbody+power law model, 

\item the components of the compound spectrum are decaying at rates
that differ significantly, and

\item the component temperatures of the blackbody emission are nearly
constant, implying that the area of emission is steadily decreasing.

\end{itemize}

Several outstanding questions remain that can be addressed by
continuing X-ray observations of \taxp\ over the next few years. Will
the current spectrum evolve into the prior quiescent one? Is the cool
quiescent spectrum still present so that it will begin to dominate
over the warm temperature component?  To what degree is the quiescent
emission pulsed? Why do the temperatures hardly change with time,
requiring the apparent decrease in area? What is the emission geometry
on the NS and what can we learn from it? Future observations will test
our predictions for the flux decay and pulsed fractions and allow a
better understanding of the relationship between the spectral
components and the pulse profile. Ultimately, the goal of this study
is to disentangle the processes of thermal and non-thermal emission to
which magnetars convert their energy, during outburst and quiescence.

\acknowledgements

We thank Fred Jansen and Norbert Schartel for providing the three
\xmm\ Target of Opportunity observations of \taxp. \xmm\ is an ESA
science mission with instruments and contributions directly funded by
ESA Member States and NASA. This research is supported by NASA grants
NNG05GC43G.

\end{document}